\documentclass[aps,prd,superscriptaddress,twoside,twocolumn,nofootinbib,10pt,%
showpacs,floatfix]{revtex4-1}

\usepackage{amsmath,amssymb}
\usepackage{graphicx,bm}
\usepackage{epstopdf}
\usepackage{ulem}
\usepackage[usenames]{color}
\usepackage{float}

\allowdisplaybreaks

\def\bi{\bibitem}
\def\lsim{\mathrel{\rlap{\lower3pt\hbox{\hskip1pt$\sim$}}
     \raise1pt\hbox{$<$}}} 
\def\gsim{\mathrel{\rlap{\lower3pt\hbox{\hskip1pt$\sim$}}
     \raise1pt\hbox{$>$}}} 

\begin{document}

\title{Chiral-scale effective theory including a dilatonic meson}

\author{Yan-Ling Li}
\affiliation{College of Physics, Jilin University, Changchun,
130012, China}

\author{Yong-Liang Ma}
\email{yongliangma@jlu.edu.cn}
\affiliation{Center for Theoretical Physics and College of Physics, Jilin University, Changchun,
130012, China}

\author{Mannque Rho}
\email{mannque.rho@cea.fr}
\affiliation{Institut de Physique Th\'eorique,
CEA Saclay, 91191 Gif-sur-Yvette c\'edex, France }

\date{\today}
\begin{abstract}
A scale-invariant chiral effective Lagrangian is constructed for octet pions and a dilaton figuring as Nambu-Goldstone bosons with vector mesons incorporated as hidden gauge fields. The Lagrangian is built to the next-to-leading order in chiral-scale counting without baryon fields and then to  leading order including baryons. The resulting theory is hidden scale-symmetric and local symmetric. We also discuss some possible applications of the present Lagrangian.
\end{abstract}

\maketitle

\section{Introduction}

The lightest scalar boson with the quantum numbers of vacuum in the strongly coupled theory has been an extremely evasive object in both particle physics and nuclear physics~\cite{pelaez}.  In particle physics, the scalar listed in the ``particle physics booklet" as $f_0(500)$ defies a simple and acceptable identification in QCD variables and a clear understanding of its role in a variety of hadronic processes, in a stark contrast to other light-quark mesons that are described as one-quark-one-antiquark systems. In nuclear physics, the problem is a lot more striking in a profound way~\cite{Walecka:1995mi}. This is the primary topic of this work.

The important role of a light scalar meson has been long recognized in the structure of nuclear forces and in the unified description of both finite and infinite nuclear systems. But what has been often invoked for nuclear dynamics is a local scalar meson field and it has been a long-standing puzzle how such a zero-width field makes sense when the observed scalar has a large width.  What is perhaps a lot more striking is the series of recent observations that bear on the origin of the proton mass and the state of matter at high density appropriate for the structure of massive compact stars. There are several indications in effective field theory models that combine hidden global and local symmetries (more details below) that the proton mass has a large component that remains non-vanishing when the order parameter for chiral symmetry in QCD {$\langle\bar{q}q\rangle$ } is dialed to zero~\cite{Detar:1988kn,hls-paritydoublet,Ma:2013ooa,Ma:2013ela}. This is contrary to the generally accepted paradigm for the source of  mass in the light-quark systems, namely, the Nambu mechanism anchored on the spontaneous breaking of chiral symmetry. This can be seen in a scale-invariant hidden local symmetry model where parity-doubling is injected {\it ab initio}  with a chiral-invariant baryon mass $m_0$ when density is increased so that the quark condensate {$\langle\bar{q}q\rangle$} goes toward zero (in the chiral limit)~\cite{hls-paritydoublet}. But more intriguingly, the parity-doubled state arises in a scale-invariant hidden local symmetric meson Lagrangian when dense matter is treated as skyrmion matter~\cite{Ma:2013ooa,Ma:2013ela}. Here there is no ab initio chiral invariant mass in the Lagrangian. However when density increases beyond a certain density $n_{1/2} \sim (2-3)n_0$ (where $n_0$ is the normal nuclear matter density), the skyrmions in the matter fractionize to half-skyrmions and the baryon-number-1 complex of half-skyrmions in the matter retains a sizable residual mass, comparable to the chiral invariant mass $m_0$ in the parity-doubling model, which remains non-vanishing at high density.  This is a phenomenon of an emergent symmetry linked to a topology change in the skyrmion model.  Remarkably, this phenomenon is associated with symmetries in nuclear dynamics at high density, including possible emergence of scale invariance and local flavor symmetry~\cite{PR-emergence},  both absent or hidden in QCD in the vacuum, and hints at how the proton mass can be understood.

In a recent article~\cite{LPR}, a resolution of the above so-called ``scalar meson conundrums"  in nuclear physics was suggested to be found in the notion that the $f_0(500)$ is a pseudo-Nambu-Goldstone boson (pNGB)  arising from the spontaneous breaking of scale invariance associated with a possible existence of an infrared (IR) fixed point in the running QCD coupling constant $\alpha_s$ proposed in ~\cite{Crewther:2013vea}\footnote{We should mention that there is an alternative approach that involves an IR fixed point~\cite{Golterman:2016lsd} in which the role of IR fixed point is given in terms of a ``critical"  number of flavors/number of colors  $n_f=N_f^c/N_c$  in what corresponds to the Veneziano limit. We make  a comparison between this approach and the C-T approach that we adopt in this  work in Appendix.}, intricately and inseparably combined with hidden local symmetry for the light-quark vector mesons $\rho$ and $\omega$~\cite{Harada:2003jx,yamawaki-GEB}. This scheme was then applied to the description of massive compact-star matter~\cite{PKLR}, in a compressed density regime which is extremely difficult to access with trustful nonperturbative tools, such as, e.g.,  lattice QCD.

This is the first in the series to quantify and go beyond what was obtained in \cite{LPR,PKLR} and address the above conundrums in a more systematic way based on chiral-scale perturbation theory with a Lagrangian that includes a dilaton and hidden local vector fields, both with and without baryons.

In a nutshell, the Crewther-Tunstall (hereafter referred to as C-T) proposal is to suppose that there is an IR fixed point at which the trace of the energy-momentum tensor (TEMT for short) $\theta_\mu^\mu$ vanishes in the chiral limit
\begin{eqnarray}
\theta_\mu^\mu & = & \frac{\beta(\alpha_s)}{4\alpha_s} G_{\mu\nu}^aG^{a\mu\nu} + (1+\gamma_m(\alpha_s))\sum_{q=u,d,s}m_q\bar{q}q \to 0  \nonumber
\end{eqnarray}
{\it with both chiral symmetry and scale symmetry spontaneously broken}. The scalar $f_0(500)$ is then identified as a Nambu-Goldstone boson, denoted as $\sigma$ \footnote{ Not to be confused with the fourth component of the chiral four-vector $(\overrightarrow{\pi},\sigma)$ in the two-flavor linear sigma model.  In this paper it's a chiral singlet. In chiral perturbation theory anchored on nonlinear sigma model, the effect of the iso-singlet sigma at low energy can be simulated by high-order loop effects~\cite{Oller:1997ti}. We do not go into the issue of how to reconcile the standard high-order chiral loop effects and the low-order scale-chiral effects,  which would require an involved work. }, arising from the spontaneously broken scale symmetry, its mass coming from the explicit breaking due to the quark mass and the departure of the gauge coupling $\alpha_s$ from the IR fixed point $\alpha_{\rm IR}$ (at which the beta function vanishes). The dilaton is to join the pseudo-scalar pseudo-Numbu-Goldstone bosons (pNBs) $\pi, \eta$ and $K$ to form a pNB multiplet. The power of this approach is to capture by tree order in chiral-scale perturbation expansion (to be described below) certain effects involving scalar excitations that are given at high orders in the standard chiral perturbation approach without the dilaton. This applies not only to the properties of elementary interactions as stressed in~\cite{Crewther:2013vea}, but more significantly in nuclear processes as discussed in \cite{LPR,PKLR}. Given the mass scale involved for the $\sigma$ comparable to that of kaons, it calls for three flavors extending the standard three-flavor chiral perturbation theory $\chi$PT$_3$ to  $\chi$PT$_\sigma$~\cite{Crewther:2013vea} (hereafter called C-T approach).

Whether or not the IR fixed point exists for three-flavor QCD in the matter-free space is not yet established and remains highly controversial. On the one hand, there are no lattice indications or compelling theoretical arguments for the existence of IR fixed points for $N_f \lsim 8$. There are also arguments based on low--energy theorems that the notion of $f_0(500)$ as a NG boson for scale symmetry is not tenable for QCD in the vacuum although it cannot be ruled out in medium~\cite{yamawaki-GEB}. On the other hand, none of the presently available lattice calculations can be taken as an unambiguous no-go theorem~\cite{lattice-motivated}.  Although not quite compelling,  there is even a tentative support from a stochastic numerical perturbation calculation in Pad\'e approximant  that indicates an IR fixed point for two-massless flavor quarks~\cite{Horsley:2013pra}.  What seems to be relevant to the issue at hand is the observation that in some strongly coupled theories such as in walking technicolor theories, there is a support for it in lattice simulations for large $N_f$~\cite{DeGrand:2015zxa}. In such walking theories, a dilatonic chiral perturbation theory for light Higgs boson similar to what we discuss in the present work has been formulated~\cite{Matsuzaki:2013eva} although the IR fixed point here involves  a conformal window which lies at much higher $N_f$. Now the recent proposal for an IR fixed point by ~\cite{Golterman:2016lsd} exploits for QCD the approach to a conformal window of the type figuring in dilatonic Higgs theory~\cite{Matsuzaki:2013eva}.

What interests us, which may not be crucially tied to the precise notion of IR fixed point in the real world of broken symmetries with masses, is the possibility that such an IR fixed point even if hidden in the vacuum can be generated, as suggested in \cite{LPR,PKLR}, as an emergent symmetry in dense matter~\cite{Beane:1994ds,Sasaki:2011ff,PR-emergence}. This is similar to the issue of restoring of $U_A (1)$ symmetry at high temperature even though the axial anomaly cannot be turned off~\cite{Pisarski:1983ms}. It is also similar to hidden local symmetry with the vector manifestation (VM) with the vanishing $\rho$ mass~\cite{Harada:2003jx}. The VM is not in QCD in the vacuum, but there is nothing to prevent it from emerging in dense/hot matter.

The application of the  C-T theory to dense baryonic matter was made in \cite{LPR,PKLR} in the spirit of mean-field approximations with a scale-chiral Lagrangian implemented with hidden local fields, with the quantum effects treated in Wilsonian RG decimation approach. To confront with Nature both in finite nuclei and in infinite nuclear matter, however, a systematic scale-chiral perturbation is mandatory. This has been seen in nuclear processes where the scalar did not figure importantly, for which the standard chiral perturbation approach at high order met with an impressive success. For a review on this aspect, see \cite{holtetal}. The power of the C-T theory is that this can be done also with the dilaton incorporated in hidden local symmetry.  As the first step toward that goal, we formulate in this paper the chiral-scale perturbation theory up to the next-to-leading order (NLO) for mesons and the heavy baryon expansion of the scale invariant chiral perturbation theory for baryons including the lightest scalar meson $f(500)$ by using C-T approach~\cite{Crewther:2013vea}. We also compare the C-T approach to the alternative approach proposed in Ref.~\cite{Golterman:2016lsd} (hereafter, called G-S approach). The Lagrangian constructed here involving three flavors will be very useful also for consistently treating strangeness in compact-star matter, a problem that has defied theorists' efforts since many years.

This paper is organized as follows: In Sec.~\ref{sec:SChPT}, we construct the scale-invariant chiral perturbation theory up to the NLO in chiral-scale counting. The typical features of the C-T approach are discussed. The scale-invariant effective theory of vector mesons based on hidden local symmetry is constructed to the NLO in Sec.~\ref{sec:SHLS}. We then turn to scale-invariant chiral effective theory of baryons and discuss the heavy-baryon expansion in Sec.~\ref{sec:HB}. A comparison between the C-T approach and the G-S approach is given in Appendix.

\section{Chiral-scale perturbation theory $\chi$PT$_\sigma$}

\label{sec:SChPT}

In Ref.~\cite{Crewther:2013vea}, $\chi$PT$_\sigma$ was explicitly given at the leading order in the chiral-scale counting (to be defined precisely below) and the procedure to go to the NLO in the meson sector was also discussed. The procedure of the construction of $\chi$PT$_\sigma$ is as follows : Since in $\chi$PT$_\sigma$, the NGBs are $\pi, K$ and $\sigma$, one first writes down all possible derivative terms acting on these particles and counting each derivative as chiral-scale order $O(p)$. In the same way as in the standard $\chi$PT (s$\chi$PT), the current quark mass is counted as chiral-scale order $O(p^2)$. Moreover, since the theory is constructed for $\alpha_s$ below but near the IR fixed point, one should expand the beta function $\beta(\alpha_s)$ and the quark mass anomalous dimension $\gamma_m(\alpha_s)$ around the IR fixed point $\alpha_{\rm IR}$ and counting $\Delta\alpha_s = \alpha_s - \alpha_{\rm IR}$ as chiral-scale order $O(p^2)$ since it is proportional to $m_\sigma^2$.

We find it more convenient to  construct the Lagrangian in terms of the conformal compensator $\chi$ which has a mass dimension-one and, under scale transformation $x \to \lambda^{-1} x$, transforms linearly as
\begin{eqnarray}
\chi(x) & \to & \lambda \chi(\lambda^{-1}x).
\end{eqnarray}
Under chiral transformation $SU(3)_L \times SU(3)_R$, $\chi$ is invariant.
Then, in terms of the NGB $\sigma$, the dilaton compensator $\chi$ is written as
\begin{eqnarray}
\chi(x) & = & f_\sigma e^{\sigma/f_\sigma},
\end{eqnarray}
where $f_\sigma$ is the $\sigma$ decay constant and, under scale transformation, $\sigma$ field transforms {\it nonlinearly} as
\begin{eqnarray}
\sigma(x) & \to & \sigma(\lambda^{-1}x) + f_\sigma\ln \lambda.
\end{eqnarray}
For the NGBs associated with the spontaneous breaking of chiral symmetry, we use the scale-and-mass dimension zero $U(x)$ with
\begin{eqnarray}
U(x) & = & e^{2i\pi_a T_a/f_\pi},
\end{eqnarray}
where $T_a = \lambda_a/2$ with $\lambda_a$ being the Gell-Mann matrices. Under chiral transformation,
\begin{eqnarray}
U(x) & \to & g_L U(x) g_R^\dagger
\end{eqnarray}
where $g_{L,R}\in SU(3)_{L,R}$ and under scale transformation,
\begin{eqnarray}
U(x) & \to & U(\lambda^{-1}x).
\end{eqnarray}

\subsection{$\chi$PT$_\sigma$ at the leading order}

Using the conformal compensator field given above, we write the effective Lagrangian, valid to the leading chiral-scale order $O(p^2)$, as given by Crewther and Tunstall~\cite{Crewther:2013vea}
\begin{eqnarray}
\cal L_{\chi {\rm PT}_\sigma}^{\rm LO} & = & {\cal L}_{{\chi {\rm PT}_\sigma};{\rm inv}}^{d=4} + {\cal L}_{{\chi {\rm PT}_\sigma};{\rm anom}}^{d > 4} + {\cal L}_{{\chi {\rm PT}_\sigma};{\rm mass}}^{d < 4},\label{eq:CTL}
\end{eqnarray}
with
\begin{subequations}
\begin{eqnarray}
{\cal L}_{\rm inv, LO}^{d=4} & = & c_1 \frac{f_\pi^2}{4} \left( \frac{\chi}{f_\sigma}\right)^2 {\rm Tr}\left( \partial_\mu U \partial^\mu U^\dagger \right) \nonumber\\
& &{} + \frac{1}{2} c_2 \partial_\mu \chi \partial^\mu \chi + c_3 \left( \frac{\chi}{f_\sigma}\right)^4, \label{eq:CTL40}\\
{\cal L}_{\rm anom, LO}^{d > 4} & = & (1 - c_1)\frac{f_\pi^2}{4} \left( \frac{\chi}{f_\sigma}\right)^{2+\beta^\prime} {\rm Tr}\left( \partial_\mu U \partial^\mu U^\dagger \right)\nonumber\\
& &{} + \frac{1}{2}(1 - c_2) \left( \frac{\chi}{f_\sigma}\right)^{\beta^\prime} \partial_\mu \chi \partial^\mu \chi \nonumber\\
& &{} + c_4 \left( \frac{\chi}{f_\sigma}\right)^{4+\beta^\prime},\label{eq:CTLg40}\\
{\cal L}_{\rm mass, LO}^{d < 4} & = &{} \frac{f_\pi^2}{4} \left( \frac{\chi}{f_\sigma}\right)^{3-\gamma_m} {\rm Tr}\left( \mathcal{M}^\dagger U + U^\dagger \mathcal{M} \right),\label{eq:CTLm40}
\end{eqnarray}
\end{subequations}
where $\mathcal{M}$ is the current quark matrix with $\mathcal{M} = {\rm diag}(m_\pi^2,m_\pi^2, 2m_K^2 - m_\pi^2)$,  $c_i$'s are unknown constants of scale-chiral order  $O(p^0)$ for $i=1,2$ and $O(p^2)$ for $i=3,4$ but mass dimension-zero for $i=1,2$ and mass dimension-four for $i=3,4$. It is important to note that the coefficients $c_3$ and $c_4$ are both implicitly $O(p^2)$ in chiral-scale counting. This is because they figure in giving mass to the dilaton similarly to $\mathcal{M}$ for the pseudo-scalar NG mesons. However differently from $\mathcal{M}$, setting to zero of which corresponds to turning off explicit chiral symmetry breaking, i.e., going to the chiral limit, setting $c_3=0$ {only} does not turn off the explicit scale symmetry breaking: It is $\beta^\prime\to 0$ that does the job. This difference should be kept in mind in keeping track of chiral-scale order and explicit symmetry breaking for the dilaton. {We will return to this matter at the end of this subsection.}

In ~\eqref{eq:CTL}, both the anomalous dimension $\gamma_m$ of the quark mass and $\beta^\prime=\partial \beta (\alpha_s)/\partial\alpha_s$ are evaluated at the IR fixed point $\alpha_{\rm IR}$. Eq.~\eqref{eq:CTL40} is the invariant term under scale transformation. \eqref{eq:CTLm40} is the current quark mass term including the anomalous dimension {of quark mass operator}. Eq.\eqref{eq:CTLg40}  accounts for the effect of the anomalous dimension of $G^2$ which depends on the slope of the beta function.

Let us briefly review the structure of the LO Lagrangian \eqref{eq:CTL}. Consider the (matter-free) vacuum in the chiral limit. In this case, since the vacuum should be stable in the $\sigma = 0$ direction, or, equivalently, $\chi = f_\sigma$, one has
\begin{eqnarray}
4c_3 + (4 + \beta^\prime)c_4 & = & 0.
\end{eqnarray}
The nontrivial solution of this equation with the lower-bound dilaton potential is
\begin{eqnarray}
c_3 & = & {} (4 + \beta^\prime)c; \quad c_4 ={} - 4c ,
\end{eqnarray}
with a positive  $c \sim O(p^2)$. By using this solution, the dilaton potential becomes
\begin{eqnarray}
V(\chi) & = & {} - (4 + \beta^\prime)c\left( \frac{\chi}{f_\sigma}\right)^4 + 4 c \left( \frac{\chi}{f_\sigma}\right)^{4 + \beta^\prime},
\label{eq:potentialchictLO}
\end{eqnarray}
which explicitly shows that, with $\beta^\prime \neq 0$, the dilaton potential is in the Nambu-Goldstone mode, i.e., the minima of the potential appears at $\langle \chi \rangle = f_\sigma$. However, if $\beta^\prime = 0$, $V(\chi) = 0$ so the scale symmetry cannot be broken. This simple observation illustrates that the spontaneous breaking of scale symmetry and explicit breaking of scale symmetry are correlated and the spontaneous breaking is triggered by explicit breaking which agrees with that unlike chiral symmetry, spontaneous breaking of scale symmetry cannot take place in the absence of explicit symmetry breaking~\cite{Freund:1969hh}. {\it This implies  that it is not possible to ``sit on" the IR fixed point with both scale and chiral symmetries spontaneously broken.} This is analogous to that one cannot ``sit" on the VM fixed point in hidden local symmetry (HLS) theory~\cite{Harada:2003jx}.  In what follows, {\it we shall therefore always consider the state of matter near but not on the IR fixed point.}

In the case that $\beta^\prime(\alpha_{\rm IR})$ is a small quantity, i.e., $\beta^\prime(\alpha_{\rm IR}) \ll 1$,  the potential~\eqref{eq:potentialchictLO} can be expanded to the leading order in $\beta^\prime$~\footnote{ Here and in what follows, we are assuming that $\beta^\prime$ can be taken as a ``small" parameter that can be used in expansion. There is, however,  an intriguing indication~\cite{Ma-Rho2016} from dense baryonic matter simulated as skyrmions on a crystal lattice that $\beta^\prime\sim (2-3)$. Obtained in a highly correlated baryonic matter, it may not be taken as a QCD quantity in the matter-free vacuum. However it could indicate that a naive expansion may not be valid in $\chi$PT$_\sigma$ applied to dense matter.}
\begin{eqnarray}
V(\chi) & = & {} \frac{m_\sigma^2 f_\sigma^2}{4} \left( \frac{\chi}{f_\sigma}\right)^4 \left[\ln \left( \frac{\chi}{f_\sigma}\right) - \frac{1}{4}\right],
\label{eq:potentiallog}
\end{eqnarray}
which is the familiar Coleman-Weinberg type potential used in the literature~\cite{Goldberger:2008zz}. In the derivation of~\eqref{eq:potentiallog}, we have used that, in the chiral limit, $m_\sigma^2 f_\sigma^2 = 4\beta^\prime(4 + \beta^\prime)c \simeq 16 \beta^\prime c$ for a small $\beta^\prime$, which, with $\beta^\prime c\sim O(p^2)$, is the dilaton analog to the Gell-Mann-Oakes-Renner relation for the pion with $m_\pi^2\sim O(p^2)$~\cite{Crewther:2013vea}.

Moreover, when $\beta^\prime(\alpha_{\rm IR})$ is a small parameter, in the chiral limit, one can rewrite the sum of ${\cal L}_{\rm inv, LO}^{d = 4}$ and ${\cal L}_{\rm anom, LO}^{d > 4}$ as
\begin{eqnarray}
{\cal L}_{{\rm LO}}^{\chi{\rm limit}}& = & {\cal L}_{\rm inv, LO}^{d = 4} + {\cal L}_{\rm anom, LO}^{d > 4} \nonumber\\
& = & \frac{f_\pi^2}{4} \left( \frac{\chi}{f_\sigma}\right)^2 {\rm Tr}\left( \partial_\mu U \partial^\mu U^\dagger \right) + \frac{1}{2} \partial_\mu \chi \partial^\mu \chi \nonumber\\
& &{} + \left[(1 - c_1)\frac{f_\pi^2}{4} \left( \frac{\chi}{f_\sigma}\right)^{2} {\rm Tr}\left( \partial_\mu U \partial^\mu U^\dagger \right) \right.\nonumber\\
& &\left.\quad \;\;{} + (1 - c_2)\frac{1}{2} \left( \frac{\chi}{f_\sigma}\right)^{2} \partial_\mu \chi \partial^\mu \chi \right]\sum_{n=1}^\infty\left(\beta^\prime \Sigma\right)^n, \nonumber\\
& & +  c_3\left( \frac{\chi}{f_\sigma}\right)^4 \left[ 1 + \frac{c_4}{c_3}\sum_{n=1}^\infty\left(\beta^\prime \Sigma\right)^n\right],
\label{eq:Lag4l4sum}
\end{eqnarray}
where $\Sigma = \ln (\chi/f_\sigma) = \sigma/f_\sigma$.
The Lagrangian ${\cal L}_{{\rm LO}}^{\chi{\rm limit}}$ explicitly shows the effect of $\beta^\prime$ which accounts for the explicit breaking of scale symmetry in the chiral limit. When scale symmetry is restored by setting $\beta^\prime = 0$~\footnote{Probably this statement makes sense only as a formal mathematical argument but cannot be done in QCD in the vacuum since there is no sense Wigner mode can be realized with scale symmetry.}, the surviving terms are scale invariant.

Next, we consider the infinitesimal scale transformation $x \to \lambda^{-1}x \equiv (1 + \theta)^{-1}x$ with $\theta$ being an infinitesimal parameter. Under such transformation, the fields in Lagrangian \eqref{eq:CTL} transform as
\begin{eqnarray}
\delta_\theta U(x) & = & {}\theta x_\mu \partial^\mu U(x) , \nonumber\\
\delta_\theta \chi(x) & = & {}\theta\left( 1 + x_\mu \partial^\mu \right) \chi(x) .
\end{eqnarray}
From these transformation rules, one can derive the following partially conserved dilatation current (PCDC) relation
\begin{eqnarray}
\left\langle \theta_\mu^\mu\right\rangle & = & \left\langle \partial_\mu D^\mu \right\rangle = {} 4 \beta^\prime c = \frac{m_\sigma^2 f_\sigma^2}{4},
\end{eqnarray}
which, in analogy to PCAC, is partially conserved dilatation current (PCDC).

Let us pause briefly to make a side remark highly pertinent to the future application to nuclear physics of what we shall develop below. In a medium of baryonic matter as discussed in \cite{LPR,PKLR}, the vacuum is modified by density.  Hence we expect that  the decay constant of $\sigma$  deviates from its vacuum value, i.e., $f_\sigma^\ast \neq f_\sigma = \langle\chi\rangle$. In \cite{LPR}, it was argued that $f_\pi^\ast = f_\pi \langle\chi\rangle$, which is consistent with the suggestion of \cite{Crewther:2013vea} that chiral symmetry breaking is locked to scale symmetry $f_\pi\approx f_\sigma$.

	

At the tree order with the Lagrangian~\eqref{eq:CTL}, apart from the phenomenological values of masses and constants, there are seven parameters $f_\sigma, \beta^\prime, \gamma_m$ and $c_i (i = 1, \cdots, 4)$. The $\sigma$ meson decay constant $f_\sigma$ is roughly estimated to be $100$~MeV in Ref.~\cite{Crewther:2013vea}. Through the minimal condition of the potential and the mass of $\sigma$, two relations among $\beta^\prime, \gamma_m, c_3$ and $c_4$ can be established. By using the $\sigma \to \pi\pi$ decay width, one can obtain a relation among $\beta^\prime, \gamma_m$ and $c_1$. Consequently, we are left with three unknown constants, for example, $\beta^\prime, \gamma_m$ and $c_2$ relating to multi-$\sigma$ interaction. It should be feasible to fix these parameters by a comprehensive analysis of hadron and nuclear phenomena that involve the scalar effect in the nucleon-nucleon interactions.

\subsection{$\chi$PT$_\sigma$ at the-next-to-leading order}

To the leading order, both $\beta^\prime$ and $\gamma_m$ are evaluated at the IR fixed point $\alpha_{\rm IR}$ which are the leading order terms in the expansions of these functions around $\Delta \alpha_s$. Therefore, the NLO Lagrangian includes, in addition to the higher
chiral-scale order corrections due to the current quark mass and derivatives on the NGBs, the leading terms in  the expansion of $\beta^\prime$ and $\gamma_m$ in $\Delta\alpha_s$. Take for instance the anomalous dimension of the mass operator. After expanding $\gamma_m$ with respect to $\alpha_{\rm IR}$, one should make the  following substitution:
\begin{eqnarray}
\chi^{\gamma_m(\alpha_{\rm IR})} & \to & \chi^{\gamma_m(\alpha_{\rm IR})}\left[1 + \sum_{n=1}^\infty C_n \left(\Delta\alpha_s \Sigma\right)^n\right],\nonumber\\
\label{eq:alphaExpan}
\end{eqnarray}
where $C_n$ are constants evaluated at $\alpha_{\rm IR}$. And the same applies to $\chi^{\beta^\prime(\alpha_{\rm IR})}$.

In the C-T approach, from the $\sigma$-to-vacuum matrix element of the divergence of the dilatation current, one can conclude the power counting
\begin{eqnarray}
\Delta \alpha_s \sim O(p^2).\label{Delta-alpha}
\end{eqnarray}
Therefore, in the effective theory, in terms of the degrees of freedom considered in the present section, the effect of $\Delta \alpha_s$ is represented by the following {chiral-scale dimension-two chiral invariant} operators:
\begin{eqnarray}
& & {\rm Tr}\left( \partial_\nu U \partial^\nu U^\dagger \right); \;\; \partial_\nu \chi \partial^\nu \chi ; \;\; {\rm Tr}\left( \mathcal{M}^\dagger U + U^\dagger \mathcal{M} \right).
\label{eq:repdeltAChpt}
\end{eqnarray}
Since the LO terms are already chiral and Lorentz invariant, the terms listed above responsible for $\Delta\alpha_s$ should be chiral and Lorentz invariant. In addition, because of the factor $\Sigma$ in the expansion~\eqref{eq:alphaExpan}, each contribution from~\eqref{eq:repdeltAChpt} is accompanied by a $\sigma$ field.

It should be noted that, even though in the expansion \eqref{eq:alphaExpan}, the term proportional to $\Sigma$  includes only one unknown parameter combination $C_1 \Delta\alpha_s$ in the microscopic theory, in the effective theory expressed in terms of the macroscopic degrees of freedom, the contribution of $C_1 \Delta\alpha_s$ is expressed in terms of a combination of the quantities in \eqref{eq:repdeltAChpt} and there is,  in front of each term,  an unknown low-energy constant that  cannot be fixed by using only the symmetry argument. We should emphasize that the contribution from all the quantities in Eq.~\eqref{eq:repdeltAChpt} get smaller as the coupling constant $\alpha_s$ runs nonperturbatively closer to the IR fixed point and vanishes at the IR fixed point. This is because,  for on-shell processes, the effect of the derivative on the NGBs translates into the momentum of the external NGBs,  therefore the mass of the NGBs, which goes toward zero as the IR fixed point is approached.

Based on what is stated above, one can straightforwardly write down the NLO Lagrangian ${\cal L}_{\chi {\rm PT}_\sigma}^{\rm  NLO}$. It includes two sectors: one, ${\cal L}_{\chi {\rm PT}_\sigma}^{{\rm LO}\times \Delta\alpha_s}$, from the $\Delta\alpha_s$ expansion of the leading order Lagrangian ${\cal L}_{\chi {\rm PT}_\sigma}^{{\rm  LO}; d> 4}$ and ${\cal L}_{\chi {\rm PT}_\sigma}^{{\rm LO}; d< 4}$ and the other one, ${\cal L}_{\chi {\rm PT}_\sigma}^{ O(p^4)}$, from the $O(p^4)$ of counting from the standard effective theory construction:
\begin{eqnarray}
{\cal L}_{\chi {\rm PT}_\sigma}^{\rm NLO} & = & {\cal L}_{\chi {\rm PT}_\sigma}^{{\rm LO}\times \Delta\alpha_s} + {\cal L}_{\chi {\rm PT}_\sigma}^{O(p^4)}.\label{thisformula}
\end{eqnarray}
For ${\cal L}_{\chi {\rm PT}_\sigma}^{{\rm LO}\times \Delta\alpha_s}$, one can write it as
\begin{widetext}
\begin{eqnarray}
{\cal L}_{\chi {\rm PT}_\sigma}^{{\rm LO}\times \Delta\alpha_s} & = & \left[(1-c_1)\frac{f_\pi^2}{4}\left(\frac{\chi}{f_\sigma}\right)^2{\rm Tr}\left( \partial_\mu U \partial^\mu U^\dagger \right) + \frac{1}{2}(1-c_2)\partial_\mu \chi \partial^\mu \chi + c_4 \left(\frac{\chi}{f_\sigma}\right)^4\right] \left(\frac{\chi}{f_\sigma}\right)^{\beta^\prime}\Sigma \nonumber\\
& & \times \left[{} D_1 {\rm Tr}\left( \partial_\nu U \partial^\nu U^\dagger \right) + D_2 \, \partial_\nu \left(\frac{\chi}{f_\sigma}\right) \partial^\nu \left(\frac{\chi}{f_\sigma}\right) + D_3\left(\frac{\chi}{f_\sigma}\right)^{1-\gamma_m}{\rm Tr}\left( \mathcal{M}^\dagger U + U^\dagger \mathcal{M} \right) \right] \nonumber\\
& &{} + {\rm Tr}\left( \mathcal{M}^\dagger U + U^\dagger \mathcal{M} \right)\left(\frac{\chi}{f_\sigma}\right)^{3-\gamma_m}\Sigma\nonumber\\
& &{} \times \left[{} D_{4} {\rm Tr}\left( \partial_\nu U \partial^\nu U^\dagger \right) + D_{5} \, \partial_\nu \left(\frac{\chi}{f_\sigma}\right) \partial^\nu \left(\frac{\chi}{f_\sigma}\right) + D_{6}\left(\frac{\chi}{f_\sigma}\right)^{1-\gamma_m}{\rm Tr}\left( \mathcal{M}^\dagger U + U^\dagger \mathcal{M} \right) \right].
\label{eq:CTChPTNLO1}
\end{eqnarray}
\end{widetext}
Note that the Lagrangian~\eqref{eq:CTChPTNLO1}, due to the factor $\ln (\chi/f_\sigma)$, does not transform homogeneously under scale transformation. This Lagrangian serves as renormalization counter terms in the loop expansion of $\chi$PT$_\sigma$~\cite{Crewther:2013vea}. This term vanishes when the explicit breaking  of scale symmetry is absent. Unlike chiral symmetry, it cannot be ``mechanically" turned off  by, say, dialling $\beta^\prime$ to zero in the Lagrangian~\eqref{eq:CTChPTNLO1}. When the explicit symmetry breaking is absent, it amounts to setting $\Sigma=\ln(\chi/f_\sigma)$  to zero, a feature that indicates that scale symmetry cannot be spontaneously broken in the absence of explicit symmetry breaking. This reflects the unfamiliar separation of roles for the chiral order counting in the coefficients $c_{3,4}$ and for the explicit symmetry breaking in $\beta^\prime$ that carries no order counting. {It is easier to see the effect of $\beta^\prime$ if we go back to the Lagrangian~\eqref{eq:Lag4l4sum} which explicitly shows that, when the scale symmetry is restored with $\beta^\prime = 0$, what's left are scale-invariant, so ${\cal L}_{\chi {\rm PT}_\sigma}^{{\rm LO}\times \Delta\alpha_s}$ is no longer present. }

As for ${\cal L}_{\chi {\rm PT}_\sigma}^{O(p^4)}$, similarly to the LO, it can be divided into the scale-invariant term ${\cal L}_{\chi {\rm PT}_\sigma; {\rm inv}}^{O(p^4); d = 4}$ , the trace anomaly term ${\cal L}_{\chi {\rm PT}_\sigma; {\rm anom}}^{O(p^4);d > 4}$ and the explicit chiral symmetry breaking term ${\cal L}_{\chi {\rm PT}_\sigma; {\rm mass}}^{O(p^4);d < 4}$. That is
\begin{eqnarray}
{\cal L}_{\chi {\rm PT}_\sigma}^{O(p^4)} & = & {\cal L}_{\chi {\rm PT}_\sigma; {\rm inv}}^{O(p^4); d = 4} + {\cal L}_{\chi {\rm PT}_\sigma; {\rm anom}}^{O(p^4);d > 4} + {\cal L}_{\chi {\rm PT}_\sigma; {\rm mass}}^{O(p^4);d < 4},
\end{eqnarray}
where
\begin{widetext}
\begin{subequations}
\begin{eqnarray}
{\cal L}_{\chi {\rm PT}_\sigma; {\rm inv}}^{O(p^4); d = 4} & = & L_1\left[{\rm Tr}\left(\partial_\mu U^\dagger \partial^\mu U\right)\right]^2 + L_2{\rm Tr}\left(\partial_\mu U^\dagger \partial_\nu U\right){\rm Tr}\left(\partial^\mu U^\dagger \partial^\nu U\right) + L_3{\rm Tr}\left(\partial_\mu U^\dagger \partial^\mu U\partial_\nu U^\dagger \partial^\nu U\right) \nonumber\\
& &{} + J_1 \partial_\nu \left(\frac{\chi}{f_\sigma}\right) \partial^\nu \left(\frac{\chi}{f_\sigma}\right) {\rm Tr}\left( \partial_\mu U \partial^\mu U^\dagger \right) + J_2 \partial_\mu \left(\frac{\chi}{f_\sigma}\right) \partial^\nu \left(\frac{\chi}{f_\sigma}\right) {\rm Tr}\left( \partial_\nu U \partial^\mu U^\dagger \right) \nonumber\\
& & {} + J_3\partial_\mu \left(\frac{\chi}{f_\sigma}\right) \partial^\mu \left(\frac{\chi}{f_\sigma}\right)\partial_\nu \left(\frac{\chi}{f_\sigma}\right) \partial^\nu \left(\frac{\chi}{f_\sigma}\right), \label{eq:CTsChPTNL4} \\
{\cal L}_{\chi {\rm PT}_\sigma; {\rm anom}}^{O(p^4);d > 4} & = & \Bigg\{\left( 1 - L_1\right)\left[{\rm Tr}\left(\partial_\mu U^\dagger \partial^\mu U\right)\right]^2 + \left(1- L_2\right){\rm Tr}\left(\partial_\mu U^\dagger \partial_\nu U\right){\rm Tr}\left(\partial^\mu U^\dagger \partial^\nu U\right) \nonumber\\
& &\;\;{} + \left(1 - L_3\right){\rm Tr}\left(\partial_\mu U^\dagger \partial^\mu U\partial_\nu U^\dagger \partial^\nu U\right)\nonumber\\
& &\;\;{} + \left(1 - J_1\right) \partial_\nu \left(\frac{\chi}{f_\sigma}\right) \partial^\nu \left(\frac{\chi}{f_\sigma}\right) {\rm Tr}\left( \partial_\mu U \partial^\mu U^\dagger \right) + \left(1 - J_2\right) \partial_\mu \left(\frac{\chi}{f_\sigma}\right) \partial^\nu \left(\frac{\chi}{f_\sigma}\right) {\rm Tr}\left( \partial_\nu U \partial^\mu U^\dagger \right) \nonumber\\
& &\;\;{} + \left(1 - J_3\right) \partial_\mu \left(\frac{\chi}{f_\sigma}\right) \partial^\mu \left(\frac{\chi}{f_\sigma}\right)\partial_\nu \left(\frac{\chi}{f_\sigma}\right) \partial^\nu \left(\frac{\chi}{f_\sigma}\right) \Bigg\} \left(\frac{\chi}{f_\sigma}\right)^{\beta^\prime},\label{eq:CTsChPTNLT4}\\
{\cal L}_{\chi {\rm PT}_\sigma; {\rm mass}}^{O(p^4);d < 4} & = & L_4 \left(\frac{\chi}{f_\sigma}\right)^{1-\gamma_m}{\rm Tr}\left(\partial_\mu U^\dagger \partial^\mu U\right){\rm Tr}\left(\mathcal{M}^\dagger U + U^\dagger \mathcal{M} \right) + L_5 \left(\frac{\chi}{f_\sigma}\right)^{1-\gamma_m}{\rm Tr}\left[\partial_\mu U^\dagger \partial^\mu U\left(\mathcal{M}^\dagger U + U^\dagger \mathcal{M} \right)\right] \nonumber\\
& &{} + L_6 \left(\frac{\chi}{f_\sigma}\right)^{2(3-\gamma_m)}\left[{\rm Tr}\left(\mathcal{M}^\dagger U + U^\dagger \mathcal{M} \right)\right]^2 + L_7 \left(\frac{\chi}{f_\sigma}\right)^{2(3-\gamma_m)}\left[{\rm Tr}\left(\mathcal{M}^\dagger U - U^\dagger \mathcal{M} \right)\right]^2 \nonumber\\
& &{}  + L_8 \left(\frac{\chi}{f_\sigma}\right)^{2(3-\gamma_m)}{\rm Tr}\left(\mathcal{M}^\dagger U\mathcal{M}^\dagger U + U^\dagger \mathcal{M} U^\dagger \mathcal{M} \right) + H_2 \left(\frac{\chi}{f_\sigma}\right)^{2(3-\gamma_m)}{\rm Tr}\left(\mathcal{M}^\dagger \mathcal{M}\right)\nonumber\\
& &{} + J_4 \left(\frac{\chi}{f_\sigma}\right)^{1-\gamma_m}\partial_\mu \left(\frac{\chi}{f_\sigma}\right) \partial^\mu \left(\frac{\chi}{f_\sigma}\right) {\rm Tr}\left(\mathcal{M}^\dagger U + U^\dagger \mathcal{M} \right).\label{eq:CTsChPTNLM4}
\end{eqnarray}
\end{subequations}
\end{widetext}
For $\beta^\prime \ll 1$ that we assume,  $\chi^{\beta^\prime}$ can be expanded and only the leading term will need to be taken for applications to nuclear physics.

In the construction given above, we did not include the external fields needed for the electroweak processes. This can be done by suitably gauging the global chiral symmetry
\begin{eqnarray}
\partial_\mu U & \to & D_\mu U = \partial_\mu U - i \mathcal{L}_\mu U + iU \mathcal{R}_\mu,
\end{eqnarray}
{
where $\mathcal{L}_\mu$ and $\mathcal{R}_\mu$ are the gauge fields of the left- and right-handed chiral transformations, respectively.
And, if we are only interested in the electromagnetic force we can choose $\mathcal{V}_\mu = \mathcal{L}_\mu + \mathcal{R}_\mu = {} -2 e Q A_\mu, \mathcal{A}_\mu = \mathcal{L}_\mu - \mathcal{R}_\mu = 0$ with $Q = {\rm diag} (2/3,-1/3,-1/3)$ being the quark charge matrix.} In the C-T theory, the chiral counting for the external fields is the same as that in s$\chi$PT.

Before closing this subsection, we list the low energy constants of $\chi$PT$_\sigma$ at NLO. In Lagrangian~\eqref{thisformula}, the low energy constants $L_i (i = 1, \cdots, 8)$ and $H_2$ take the same values as that in $\chi$PT$_3$ {since, when the dilaton field is switched-off, $\chi = \langle \chi\rangle = f_\sigma$}. In addition to $L_i (i = 1, \cdots, 8)$ and $H_2$, we have twelve more parameters $D_i (i = 1, \cdots, 8)$ and $J_i (i = 1, \cdots, 4)$.

\section{Scale-invariant hidden local symmetry: HLS$_{\sigma}$ }

\label{sec:SHLS}

We next extend the above discussion of $\chi$PT$_\sigma$ to chiral effective theory including the lowest-lying vector mesons via  the hidden local symmetry strategy~\cite{Bando:1984ej,Bando:1987br,Harada:2003jx}, HLS$_{\sigma}$. Here we use the conformal compensator in the HLS Lagrangian of \cite{Harada:2003jx} to make it gauge equivalent to $\chi$PT$_\sigma$ after integrating out the vector mesons. We are employing $[SU(3)]_{\rm HLS}$ for the hidden local symmetry. The extension to other HLS, such as $[U(3)]_{\rm HLS}$ or $[SU(3)\times U(1)]_{\rm HLS}$, is straightforward.

In HLS, we decompose the field $U(x)$ as
\begin{equation}
U(x)=\xi_L^\dag(x)\xi_R(x).
\end{equation}
Due to the redundancy in this decomposition, one can implement a gauge symmetry $H_{\rm local}$ by requiring that $\xi_{L,R}$ have the transformation
\begin{equation}
\xi_{L,R}(x)\mapsto\xi_{L,R}^\prime(x) =
h(x)\xi_{L,R}(x)g_{L,R}^\dag,
\label{hidden-trans}
\end{equation}
where $h(x) \in H_{\rm local}$.

Corresponding to the HLS gauge symmetry $[{\rm SU}(3)]_{\rm HLS}$ one can introduce the gauge
field $V_\mu(x)$ that transforms as
\begin{eqnarray}
V_\mu(x) \rightarrow V_\mu(x)^\prime & = &h(x)V_\mu(x) h^\dag(x) - i\partial_\mu h(x) \cdot h^\dag(x),
\nonumber
\end{eqnarray}
where, now, $h(x) \in [{\rm SU}(3)]_{\rm HLS}$.
With  the chiral fields $\xi_{L,R}$ and the hidden gauge bosons $V_\mu$
one can define the following two Maurer-Cartan $1$-forms:
\begin{eqnarray}
\hat{\alpha}_{\parallel\mu} & = & \frac{1}{2i}(D_\mu \xi_R \cdot
\xi_R^\dag +
D_\mu \xi_L \cdot \xi_L^\dag), \nonumber\\
\hat{\alpha}_{\perp\mu} & = & \frac{1}{2i}(D_\mu \xi_R \cdot
\xi_R^\dag - D_\mu \xi_L \cdot \xi_L^\dag)
\ , \label{eq:1form}
\end{eqnarray}
where the covariant derivative is defined as $D_\mu \xi_{R,L} = (\partial_\mu - i V_\mu)\xi_{R,L}$.  Both of these quantities transform as $\hat{a}_{\parallel,\perp}^{\mu} \rightarrow h(x) \hat{a}_{\parallel,\perp}^{\mu} h(x)^\dag$.
For the gauge field $V_\mu$ we have the field strength tensor
\begin{eqnarray}
V_{\mu\nu}(x) = \partial_\mu V_\nu(x)-\partial_\nu
V_\mu(x)-i[V_\mu(x),V_\nu(x)],\label{eq:hiddentensor}
\end{eqnarray}
with the transformation property $V_{\mu\nu}(x) \rightarrow h(x) V_{\mu\nu}(x) h(x)^\dag$.

The HLS Lagrangian are constructed in terms of the independent quantities
\begin{eqnarray}
\hat{\alpha}_{\parallel\mu},\quad \hat{\alpha}_{\parallel\mu},\quad \frac{1}{g}V_{\mu\nu},\quad {\hat{\mathcal{M}},}
\end{eqnarray}
where $g$ is the gauge coupling constant of HLS {and $\hat{\mathcal{M}} \equiv \xi_L \mathcal{M} \xi_R^\dagger$ transforming under HLS as $\hat{\mathcal{M}} \to h(x) \hat{\mathcal{M}} h^\dagger(x)$.}. In this equation, all the quantities are $O(p)$ of the chiral counting.

Now to generate the masses of the gauge bosons, we use the Higgs mechanism taking the unitary gauge
\begin{eqnarray}
\xi_L^\dag & = & \xi_R \equiv \xi = e^{i\pi/f_\pi} . \label{eq:hiddenbreaking}
\end{eqnarray}
After the gauge fixing, the massive vector meson fields are assigned to the HLS through
\begin{eqnarray}
V_\mu & = & \frac{g}{\sqrt{2}}\left(
                         \begin{array}{ccc}
                           \frac{1}{\sqrt{2}}(\rho_\mu^0 + \omega_\mu) & \rho_\mu^+ & K_\mu^{\ast,+} \\
                           \rho_\mu^- & {}-\frac{1}{\sqrt{2}}(\rho_\mu^0 - \omega_\mu) & K_\mu^{\ast,0}\\
                           K_\mu^{\ast,-} & \bar{K}_\mu^{\ast,0} & \phi_\mu \\
                         \end{array}
                       \right).
\end{eqnarray}

It is easy to incorporate the dilaton in HLS and obtain HLS$_\sigma$. As in the $\chi$PT$_\sigma$ case, the leading-order HLS Lagrangian consists of three components:
\begin{eqnarray}
\cal L_{{\rm HLS}_\sigma}^{\rm LO}& = & {\cal L}_{{\rm HLS}_\sigma;{\rm inv}}^{d=4} + {\cal L}_{{\rm HLS}_\sigma;{\rm anom}}^{d > 4} + {\cal L}_{{\rm HLS}_\sigma;{\rm mass}}^{d < 4},\label{eq:CTHLS}
\end{eqnarray}
with
\begin{widetext}
\begin{subequations}
\begin{eqnarray}
{\cal L}_{{\rm HLS}_\sigma; {\rm inv}}^{d=4} & = & f_\pi^2 {h_1} \left(\frac{\chi}{f_\sigma}\right)^2{\rm
Tr}[\hat{a}_{\perp\mu}\hat{a}_{\perp}^{\mu}] + a f_\pi^2 {h_2}  \left(\frac{\chi}{f_\sigma}\right)^2{\rm
Tr}[\hat{a}_{\parallel\mu}\hat{a}_{\parallel}^{\mu}] -
\frac{1}{2g^2}{h_3}{\rm Tr}[V_{\mu\nu}V^{\mu\nu}] + \frac{1}{2}{h_4} \partial_\mu \chi \partial^\mu \chi + {h_5} \left(\frac{\chi}{f_\sigma}\right)^4, \label{eq:CTHLS40}\\
{\cal L}_{{\rm HLS}_\sigma; {\rm anom}}^{d > 4} & = & f_\pi^2(1-{h_1}) \left(\frac{\chi}{f_\sigma}\right)^{2+\beta^\prime}{\rm
Tr}[\hat{a}_{\perp\mu}\hat{a}_{\perp}^{\mu}] + (1-{h_2}) af_\pi^2\left(\frac{\chi}{f_\sigma}\right)^{2+\beta^\prime}{\rm
Tr}[\hat{a}_{\parallel\mu}\hat{a}_{\parallel}^{\mu}]\nonumber\\
& &{}  -
\frac{1}{2g^2}(1-{h_3})\left(\frac{\chi}{f_\sigma}\right)^{\beta^\prime}{\rm Tr}[V_{\mu\nu}V^{\mu\nu}] + \frac{1}{2}(1-{h_4}) \left(\frac{\chi}{f_\sigma}\right)^{\beta^\prime} \partial_\mu \chi \partial^\mu \chi + {h_6} \left(\frac{\chi}{f_\sigma}\right)^{4+\beta^\prime}, \label{eq:CTHLSg40}\\
{\cal L}_{{\rm HLS}_\sigma; {\rm mass}}^{d < 4} & = &{} \frac{f_\pi^2}{4} \left(\frac{\chi}{f_\sigma}\right)^{3-\gamma_m}{\rm Tr}\left(\hat{\mathcal{M}} + \hat{\mathcal{M}}^\dagger\right).\label{eq:CTHLSm40}
\end{eqnarray}
\end{subequations}
\end{widetext}

To construct the NLO HLS$_\sigma$ one should include all the possible operators expressed in terms of the macroscopic degrees of freedom in the effective theory representing $\Delta\alpha_s$ in the expansion of $\gamma_m$ and $\beta^\prime$. For the HLS with $[{\rm SU}(3)]_{\rm HLS}$, the following {chiral-scale dimension-two HLS-invariant} quantities are needed:
\begin{eqnarray}
& & {\rm
Tr}(\hat{a}_{\perp\mu}\hat{a}_{\perp}^{\mu}); \qquad {\rm
Tr}(\hat{a}_{\parallel\mu}\hat{a}_{\parallel}^{\mu}); \qquad
\frac{1}{2g^2}{\rm
Tr}(V_{\mu\nu}V^{\mu\nu}); \nonumber\\
& & \partial_\mu \chi \partial^\mu \chi; \qquad {\rm Tr}\left(\hat{\mathcal{M}} + \hat{\mathcal{M}}^\dagger\right).
\end{eqnarray}
From this discussion, one can easily arrive at the NLO HLS$_\sigma$ following the approach used in $\chi$PT$_\sigma$. We will not write the explicit form here.

In the LO HLS$_\sigma$~\eqref{eq:CTHLS}, except for the parameter $a$ which comes out to be $a=2$ from the vector meson dominance, the KSRF relation and the pion decay constant $f_\pi$, there are nine parameters $f_\sigma, \beta^\prime, {\gamma_m}$ and ${h_i } (i = 1, \cdots, 6)$. In the C-T scheme, the $\sigma$ decay constant $f_\sigma$ is taken as $f_\sigma \simeq 100$~MeV. Using the three relations, the minimization of the potential, the $\sigma$ mass and the $\sigma\to \pi\pi$ decay, calculated in the $\chi$PT$_\sigma$ approach, we can reduce the number of the free parameters to five, $\beta^\prime, \gamma_m, c_2, c_3$ and $c_4$. The parameters $c_2$ and $c_3$ can be estimated or constrained with the processes that involve the vector mesons, such as $K^\ast$.

\section{Scale-invariant hidden local symmetry with baryon octet: $bs$HLS}

\label{sec:HB}

We next turn to the dilaton compensated baryonic chiral perturbation theory including the vector mesons through hidden local symmetry, $bs$HLS. In our procedure, we first write down baryonic chiral perturbation theory with HLS, denoted $b$HLS,  with the minimal number of derivatives and, then, couple the dilaton to it by means of the conformal compensator. We then make a heavy-baryon expansion of $bs$HLS to construct the dilatonic heavy-baryon chiral perturbation with hidden local symmetry, $hbs$HLS. This would give a well-defined power counting rule. In our construction, the symmetry pattern involved is  $[SU(3)_L\times SU(3)_R]_{\rm chiral} \times [SU(3)]_{\rm HLS}$. An extension to other symmetry patterns is straightforward.

In the present work, we consider the octet baryon in the ground state with quantum numbers $\frac{1}{2}^+$:
\begin{eqnarray}
B(x) & = & \left(
             \begin{array}{ccc}
               \frac{1}{\sqrt{2}}\Sigma^0 + \frac{1}{\sqrt{6}}\Lambda & \Sigma^+ & p \\
               \Sigma^- &{} - \frac{1}{\sqrt{2}}\Sigma^0 + \frac{1}{\sqrt{6}}\Lambda & n \\
               \Xi^- & \Xi^0 & {}-\frac{2}{\sqrt{6}}\Lambda \\
             \end{array}
           \right),\nonumber\\
\end{eqnarray}
Under HLS, $B(x)$ transforms as
\begin{eqnarray}
B(x) \to h(x) B(x) h^\dagger(x),
\end{eqnarray}
i.e., in the adjoint representation of $[SU(3)]_{\rm HLS}$.

In the construction, we use the canonical dimension of the baryon field, i.e., under scale transformation, it transforms
as
\begin{eqnarray}
B(x) \to \lambda^{3/2}B(\lambda^{-1}x).
\end{eqnarray}

After specifying the HLS and scale transformation properties of baryon field $B$, we are ready to write down the effective theory. Similarly to the mesonic case, the baryonic Lagrangian, at the leading order, includes the scale-invariant part and the trace anomaly part. It should be noted that, at leading order, there are no $d < 4$ terms, because the light quark mass is counted as $O(p^2)$ in $\chi$PT. The Lagrangian is written as
\begin{eqnarray}
{\cal L}_{bs{\rm HLS}} & = & {\cal L}_{bs{\rm HLS}; {\rm inv}}^{d = 4} + {\cal L}_{bs{\rm HLS}; {\rm anom}}^{d > 4},
\label{eq:LagbsHLS}
\end{eqnarray}
with
\begin{widetext}
\begin{subequations}
\begin{eqnarray}
{\cal L}_{bs{\rm HLS}; {\rm inv}}^{{\rm LO}; d = 4} & = & {g_1}{\rm Tr}\left( \bar{B} i \gamma_\mu D^\mu B\right) - \tilde{\mathring{m}}_B \frac{\chi}{f_\sigma}{\rm Tr}\left( \bar{B} B\right) - \tilde{g}_{A_1}{\rm Tr}\left( \bar{B} \gamma_\mu \gamma_5 \left\{ \hat{\alpha}_\perp^\mu,  B \right\} \right) \nonumber\\
& &{} - \tilde{g}_{A_2}{\rm Tr}\left( \bar{B} \gamma_\mu \gamma_5 \left[ \hat{\alpha}_\perp^\mu,  B \right] \right) + \tilde{g}_{V_1}{\rm Tr}\left( \bar{B} \gamma_\mu \left\{ \hat{\alpha}_\parallel^\mu,  B \right\} \right) + \tilde{g}_{V_2}{\rm Tr}\left( \bar{B} \gamma_\mu \left[ \hat{\alpha}_\parallel^\mu,  B \right] \right) \, ,\label{LBCTd4} \\
{\cal L}_{bs{\rm HLS}; {\rm anom}}^{{\rm LO}; d > 4} & = & \Big[(1-{g_1}){\rm Tr}\left( \bar{B} i \gamma_\mu D^\mu B\right) - (\mathring{m}_B - \tilde{\mathring{m}}_B) \frac{\chi}{f_\sigma} {\rm Tr}\left( \bar{B} B\right) - (g_{A_1} - \tilde{g}_{A_1}){\rm Tr}\left( \bar{B} \gamma_\mu \gamma_5 \left\{ \hat{\alpha}_\perp^\mu,  B \right\} \right)\nonumber\\
& &\;{} - (g_{A_2} - \tilde{g}_{A_2}){\rm Tr}\left( \bar{B} \gamma_\mu \gamma_5 \left[ \hat{\alpha}_\perp^\mu,  B \right] \right) + (g_{V_1} - \tilde{g}_{V_1}){\rm Tr}\left( \bar{B} \gamma_\mu \left\{ \hat{\alpha}_\parallel^\mu,  B \right\} \right) \nonumber\\
& &\;{} + (g_{V_2} - \tilde{g}_{V_2}){\rm Tr}\left( \bar{B} \gamma_\mu \left[ \hat{\alpha}_\parallel^\mu,  B \right] \right) \Big]\left(\frac{\chi}{f_\sigma}\right)^{\beta^\prime} ,
\label{LBCT2}
\end{eqnarray}
\end{subequations}
\end{widetext}
where $\mathring{m}_B$ and $\tilde{\mathring{m}}_B$ account for the baryon mass in the chiral limit, all the coupling constants are dimensionless and the covariant derivative is defined as
\begin{equation}
D^\mu B = \partial^\mu B -i \left[ V^\mu, B \right]\, .
\end{equation}
Note that, because of the conservation of the vector current of baryon, there is no term ${\rm Tr}\left( \bar{B}\gamma_\mu \partial^\mu \chi B\right)$ in Eq.~\eqref{LBCTd4}, hence  no term ${\rm Tr}\left( \bar{B}\gamma_\mu \partial^\mu \chi B\right)(\chi/f_\sigma)^\beta$ in Eq.~\eqref{LBCT2}.

In ${\cal L}_{bs{\rm HLS}}$, both $\mathring{m}_B$ and $\tilde{\mathring{m}}_B$ are taken as scale invariant quantities, so we couple the dilaton to the baryon mass term in terms of the conformal compensator to make the term $\tilde{\mathring{m}}_B (\chi/f_\sigma){\rm Tr}\left( \bar{B} B\right)$ in Eq.~\eqref{LBCTd4} scale-invariant. {It should be noted that there is an ambiguity in the meaning of the baryon mass in the chiral limit when a dilaton is present \`a la C-T. This is because in the C-T theory, the quark condensate and the dilaton condensate are locked to each other. This means that in the chiral limit, $\mathring{m}_B$ remains as is but the fluctuating dilaton field couples to the baryon as $g_{\sigma BB} \bar{\psi}\psi$.}

As is well known, the chiral counting in chiral Lagrangian including baryon is subtle because of the large baryon mass in the chiral limit. Therefore, a heavy-baryon formulation should be set up~\cite{Jenkins:1990jv,Bernard:1992qa}. To formulate the Lagrangian~\eqref{eq:LagbsHLS} in terms of heavy-baryon mass expansion, let us consider the terms contributing to the baryon mass in ~\eqref{eq:LagbsHLS}:
\begin{eqnarray}
{\cal L}_{bs{\rm HLS}}^{\rm mass} & = & \left[\tilde{\mathring{m}}_B \frac{\chi}{f_\sigma} + (\mathring{m}_B - \tilde{\mathring{m}}_B) \frac{\chi}{f_\sigma}\left(\frac{\chi}{f_\sigma}\right)^{\beta^\prime}\right]{\rm Tr}\left( \bar{B} B\right).
\nonumber\\
\end{eqnarray}
From this term, one can easily see that, after spontaneous breaking of scale symmetry, i.e., $\langle \chi\rangle = f_\sigma$, the baryon mass becomes $\mathring{m}_B$. For making heavy baryon mass $\mathring{m}_B$ expansion, we first consider a small $\beta^\prime$, such that
\begin{eqnarray}
{\cal L}_{bs{\rm HLS}}^{\rm mass} & = & \left[\mathring{m}_B + (\mathring{m}_B - \tilde{\mathring{m}}_B) \sum_{n=1}^\infty\frac{1}{n!}\left(\beta^\prime \Sigma\right)^{n}\right]\frac{\chi}{f_\sigma}{\rm Tr}\left( \bar{B} B\right).
\nonumber\\
\end{eqnarray}
We further expand the dilaton field $\chi$ around its vacuum expectation value. Then ${\cal L}_{bs{\rm HLS}}^{\rm mass}$ is expressed as
\begin{eqnarray}
{\cal L}_{bs{\rm HLS}}^{\rm mass} & = & \mathring{m}_B {\rm Tr}\left( \bar{B} B\right) + \mathring{m}_B \sum_{m=1}^\infty\frac{1}{m!}\Sigma^m {\rm Tr}\left( \bar{B} B\right)\nonumber\\
& &{} + (\mathring{m}_B - \tilde{\mathring{m}}_B) \sum_{n=1}^\infty\frac{1}{n!}\left(\beta^\prime \Sigma\right)^{n}\nonumber\\
& &\qquad\qquad{} \times \left(1+ \sum_{m=1}^\infty\frac{1}{m!}\Sigma^m\right) {\rm Tr}\left( \bar{B} B\right).
\label{eq:BMassExp}
\end{eqnarray}
In this Lagrangian, the first term is the baryon mass term in the chiral limit while the other terms are the $\sigma^n \bar{B}B \, (n = 1,2,\cdots)$ interaction terms. For example, the $\sigma$-$B$-$B$ coupling is written as
\begin{eqnarray}
g_{\sigma BB} & = & \mathring{m}_B \frac{1}{f_\sigma} + ( \mathring{m}_B- \tilde{\mathring{m}}_B ) \beta^\prime \frac{1}{f_\sigma} ,
\end{eqnarray}
which, after removing the explicit scale symmetry breaking by setting $\beta^\prime=0$, is the Goldberger-Trieman type relation~\cite{Carruthers:1971vz} in the dilaton sector in the chiral limit.

To finalize the heavy-baryon expansion, we should set up the chiral-scale counting of the interaction terms. Since the dilaton couples to baryons nonderivatively, one cannot do the usual power counting as with the derivative in pion-nucleon coupling. In the present work, in the absence of first-principle guidance, we establish the power counting using a  numerical estimation. If we take the nucleon mass in the chiral limit $\mathring{m}_B \simeq 900$~MeV~\cite{Li:2008hp}, by taking $f_\sigma \simeq f_\pi$, we obtain $g_{\sigma BB} \simeq 10$ which is close to $g_{\pi NN} = 13$~\cite{Schroder:2001rc}. This suggests that the other terms in \eqref{eq:BMassExp} could be considered as of chiral-scale order $O(p)$. That is, in terms of the compensator $\chi$
\begin{eqnarray}
& & \mathring{m}_B \left(\frac{\chi}{f_\sigma} - 1\right) + (\mathring{m}_B - \tilde{\mathring{m}}_B) \left[\left(\frac{\chi}{f_\sigma}\right)^{\beta^\prime} - 1\right] \frac{\chi}{f_\sigma} \sim O(p),
\nonumber\\
\label{eq:CountBBS}
\end{eqnarray}
in which the NGB $\sigma$ appears in the similar fashion as the pion in $U(x)$. Once the chiral-scale counting discussed here is accepted, the heavy-baryon formalism of $bs$HLS can be derived straightforwardly.

\section{Discussion and perspective}

In this paper, we constructed by means of the C-T approach the dilatonic chiral effective theories for pseudoscalar mesons, vector mesons and octet baryons by using the conformal compensator. The basic premise in this -- and forthcoming work -- is that although it may not make sense or be feasible to reach precisely the IR fixed point which would require turning off the trace anomaly, it makes sense to fluctuate close to it with the departure from it taken as a small parameter, in a way analogous to the axial anomaly or the vector manifestation fixed point of HLS. The crucial aspect of the approach is that the dilaton generically satisfies soft-dilaton theorems in a manner closely parallel to soft-pion theorems and it renders the leading-order calculations as made in \cite{LPR,PKLR} trustworthy.

The present construction of the $bs$HLS Lagrangian can be extended in a straightforward way to include the parity-doublet structure of baryons~\cite{Detar:1988kn,Jido:2001nt} with the transformation of baryons given by Refs.~\cite{Paeng:2011hy}. We relegate the details to a future publication.

The chiral-scalar effective theory discussed here can be used in the study of dense matter physics for verifying the validity of, and going beyond, the mean-field-based analysis given in~\cite{LPR,PKLR}. The possibility that scale symmetry can emerge in dense matter with $f_\sigma^\ast$ going to zero in the dilaton-limit fixed point even if $\beta^\prime\neq 0$ as discussed in \cite{PR-emergence} could be explored in this formalism in a manner similar to the two-flavor chiral perturbation approach $\chi$PT$_2$ and the RG approach based on $\chi$PT$_\sigma$ for compact-star properties as discussed in \cite{holtetal}.
In Refs.~\cite{Park:2008zg,Ma:2013ela}, it has been shown by using a chiral-scale Lagrangian with the log-type potential (\ref{eq:potentiallog}) that the dilaton suppresses the baryon mass and restores the scale symmetry at high density with $f_\sigma^\ast\to 0$. In the present construction, the explicit scale symmetry can be taken into account by the deviation from the IR fixed point $\alpha_{\rm IR}$ which gives the Lagrangian used in Refs.~\cite{Park:2008zg,Ma:2013ela} at the leading order of small $\beta^\prime$, so we believe the present Lagrangian can yield a result closer to nature. Moreover, since the present chiral-scale effective theory is constructed with three flavors, it can provide a systematic way to study effects of strangeness in nuclear matter which has hitherto not been feasible in a consistent way.

At first sight it may look like a highly daunting task to make predictions with a higher-order $\chi$PT$_\sigma$, given the exploding number of parameters in the Lagrangian, beyond the leading order in the given scale-chiral counting. However this may be too pessimistic. For instance what happens in HLS for vector mesons indicates that with an astuteness and insight one should be able to control the parameters. In HLS, if one goes beyond $O(p^2)$, the number of parameters also explodes, but it has been found that a chiral perturbation calculation with HLS can be efficiently done~\cite{Harada:2003jx} and furthermore at order $O(p^2)$, some remarkable things -- such as the KSRF relation, the vector dominance etc. -- are obtained~\cite{komargodski,Harada:2003jx}.  In a similar vein, it seems feasible to gain access to certain processes in nuclear dynamics that cannot be accessed without light scalar degrees of freedom, the most famous example being  the indispensable scalar attraction in nuclear potentials, in relativistic mean field theory etc.  How the dilaton enters in the scheme is similar to the way pseudscalar pNG mass enters in $\chi$PT$_3$ and  in the present case, more selectively in the process where the scalar quantum numbers are involved. Using the dilaton mass as the imprint of the scale symmetry explicit breaking, one should be able to do a systematic calculation of dilaton-involved nuclear processes.

In addition to nuclear dynamics under extreme conditions that we are primarily interested in, as pointed out before, the presence  of an IR fixed point with certain properties, such as low-energy theorems, analogous to what we encounter in dense baryonic matter plays an important role in strongly coupled systems which are intrinsically different from QCD matters, an intriguing recent development being certain technicolor theories purported to go beyond the Standard Model in particle physics\cite{yamawaki-GEB}.


\appendix

\section{Comparison between the C-T and G-S approaches}

We make a brief comparison between the C-T approach and an approach proposed in Ref.~\cite{Golterman:2016lsd} (G-S approach), both anchored on an IR fixed point. To the leading order in chiral-scale counting, we expect them to give the same results in nuclear physics although the mechanism leading to the IR fixed point is quite different. Whether they differ at higher orders is not obvious and will have to be worked out.

In the G-S approach, the IR fixed point involves three limits: the chiral limit, the Veneziano limit and the $n_f \to n_f^\ast$ limit with $n_f = N_f/N_c$ where $n_f^\ast$ is the lowest $n_f$ for a confined theory. The effective field theory in the scale symmetry broken phase  is expanded in terms of the three small parameters: the quark mass (pseudo-NG mass), $1/N_c$
and $\Delta n_f\equiv n_f-n_f^\ast$. All these three parameters are counted as $O(p^2)$ in the usual power counting.

The G-S Lagrangian at the leading order is~\cite{Golterman:2016lsd}
\begin{eqnarray}
{\cal L}_{\chi{\rm PT}}^{\rm LO;G-S} & = & {\cal L}_\pi^{\rm GS} + {\cal L}_\tau^{\rm GS} + {\cal L}_m^{\rm GS} + {\cal L}_d^{\rm GS}, \label{eq:GSL}
\end{eqnarray}
with
\begin{subequations}
\begin{eqnarray}
{\cal L}_\pi^{\rm GS} & = & \frac{f_\pi^2}{4}V_\pi(\sigma - \tau) \left(\frac{\chi}{f_\sigma}\right)^2 {\rm Tr}\left( \partial_\mu U \partial^\mu U^\dagger \right),\label{eq:GSLp}\\
{\cal L}_\tau^{\rm GS} & = & \frac{f_\sigma^2}{2}V_\sigma(\sigma - \tau) \left( \partial_\mu \chi \partial^\mu \chi  \right) , \label{eq:GSt}\\
{\cal L}_m^{\rm GS} & = &{} \frac{f_\pi^2}{4}V_M(\sigma - \tau) \left(\frac{\chi}{f_\sigma}\right)^{3-\gamma_m} {\rm Tr}\left( \mathcal{M}^\dagger U + hc \right), \label{eq:GSm}\\
{\cal L}_d^{\rm GS} & = & \left(\frac{\chi}{f_\sigma}\right)^4V_d(\sigma - \tau) .\label{eq:GSd}
\end{eqnarray}
\end{subequations}
Note that we have changed the notation of Ref.~\cite{Golterman:2016lsd} to that we used in the present paper. Here $V$'s are arbitrary functionals of $(\sigma-\tau)$ and $\tau$ stands for the spurion field for the source that transforms as does the dilaton $\sigma$. Since under scale transformation, the transformation of the argument $\sigma$ in $V_I, (I = \pi, \sigma, M, d)$, is cancelled by that of $\tau$, $V_I$ are scale-invariant. The trace anomaly, i.e., explicit breaking of scale symmetry, is generated when the external source is turned off.

To make a systematic expansion, one expands the potential $V_I$ terms of the argument $\sigma-\tau$
\begin{eqnarray}
V_I & = & \sum_{n=0}^\infty \frac{c_{I,n}}{n!}(\tau-\sigma)^n ,
\end{eqnarray}
and then makes explicit the dependence on  $(n_f - n_f^\ast)^n$ by expanding $c_{I,n}$ as
\begin{eqnarray}
c_{I,n} &= & \sum_{k=0}^\infty \tilde{c}_{I,nk}(n_f - n_f^\ast)^k .
\label{eq:powerCI}
\end{eqnarray}
The low-energy coefficients $c_{I,nk}$ are then endowed with the power counting~\cite{Golterman:2016lsd}
\begin{eqnarray}
c_{I,nk} & \sim & \left\{
                   \begin{array}{ll}
                     0, & \hbox{$k < n$;} \\
                     O(\delta^n), & \hbox{$k \geq n$.}
                   \end{array}
                 \right.
  ,
\label{eq:powerCI}
\end{eqnarray}
where $O(\delta)\sim O(p^2)$.
Therefore, in contrast to the ordinary chiral perturbation theory in which the low-energy constants do not carry any chiral dimension, the coefficients of the scale-invariant potentials carry power counting. This is the same in the C-T approach.

For the leading order Lagrangian, we can choose $V_\pi = V_\tau = V_M = 1$ by definition. Neglecting the source $\tau$, we then have
\begin{subequations}
\begin{eqnarray}
{\cal L}_\pi^{(\delta)} & = & \frac{f_\pi^2}{4} \left(\frac{\chi}{f_\sigma}\right)^2 {\rm Tr}\left( \partial_\mu U \partial^\mu U^\dagger \right), \label{eq:GSPLO}\\
{\cal L}_\tau^{(\delta)} & = & \frac{f_\tau^2}{2} \left( \partial_\mu \chi \partial^\mu \chi  \right) , \\
{\cal L}_m^{(\delta)} & = &{} \frac{f_\pi^2}{4} \left(\frac{\chi}{f_\sigma}\right)^{3-\gamma_m} {\rm Tr}\left( \mathcal{M}^\dagger U + U^\dagger \mathcal{M} \right), \\
{\cal L}_d^{(\delta)} & = & \left(\frac{\chi}{f_\sigma}\right)^4\left(c_{0} + c_{1}\ln \frac{\chi}{f_\sigma} \right) ,
\label{eq:dChPT2}
\end{eqnarray}
\end{subequations}
where, for convenience, we have written $\sigma$ as $\ln \chi/f_\sigma$.

By using the saddle-point condition, the dilaton potential with lower bound is written as
\begin{eqnarray}
\tilde{\cal L}_d^{(\delta)} & = &{} - c \left(\frac{\chi}{f_\sigma}\right)^4\left(\ln \frac{\chi}{f_\sigma} - \frac{1}{4}\right) .
\label{eq:LdGS}
\end{eqnarray}
Since the positive parameter $c$ is related to the dilaton mass as  $c = m_\sigma^2 f_\sigma^2/4$, one concludes that the dilaton potential \eqref{eq:LdGS} in the G-S approach is that same as \eqref{eq:potentiallog} in the C-T approach. In addition, the relation $c = m_\sigma^2 f_\sigma^2/4$ indicates that the coefficient of the dilaton potential in the G-S approach is of $O(p^2)$ which is the same as in the C-T approach up to $O(\beta^\prime)$.

From~\eqref{eq:GSPLO} we find that in the G-S approach in the chiral limit, the $\sigma\pi\pi$ coupling is given by
\begin{eqnarray}
{\cal L}_{\sigma\pi\pi}^{G-S} & = & \frac{1}{f_\sigma}\sigma |\partial \bm{\pi}|^2,
\end{eqnarray}
which yields $g_{\sigma\pi\pi} ={} - m_\sigma^2/f_\sigma$ for an on-shell dilaton. On the other hand, in the C-T approach, $g_{\sigma\pi\pi} ={} - (2 + (1-c_1)\beta^\prime)m_\sigma^2/(2f_\sigma)$.
Thus there is a slight difference in how the explicit symmetry breaking figures in the $\sigma$ coupling. This will apply at higher orders -- and multi-dilaton interactions -- with the $\sigma$ entering in the Feynman diagrams.

\acknowledgments

\label{ACK}

One of the authors (MR) would like to thank Rod Crewther, Lewis Tunstall and Koichi Yamawaki for numerous discussions by correspondence on scale symmetry in particle physics. None of them should be held responsible for possible flaws in our application of the subtle notion of scale invariance to nuclear physics. This work was supported in part by National Science Foundation of China (NSFC) under Grant No. 11475071, 11547308 and the Seeds Funding of Jilin University.

\end{document}